\documentclass[]{aa}

\usepackage{graphicx}

\usepackage{txfonts}

\usepackage[]{hyperref}
\usepackage{caption}
\usepackage{subcaption}
\usepackage{gensymb}
\usepackage[version=4]{mhchem} 

\usepackage[dvipsnames]{xcolor}

\begin{document}

   \title{Inner Disc and Circumplanetary Material in the PDS 70 System: Insights from Multi-Epoch, Multi-Frequency ALMA Observations}

   \author{D.~Fasano
           \inst{\ref{inst1},\ref{inst2}}
          \and
          M.~Benisty
          \inst{\ref{inst2}}
          \and
          P.~Curone
          \inst{\ref{inst3}}
          \and
          S.~Facchini
          \inst{\ref{inst4}}
          \and
          F.~Zagaria
          \inst{\ref{inst2}}
          \and
          T. C.~Yoshida
          \inst{\ref{inst5},\ref{inst6}}
          \and
          K.~Doi
          \inst{\ref{inst2}}
          \and
          A.~Sierra
          \inst{\ref{inst7}}
          \and
          S.~Andrews
          \inst{\ref{inst8}}
          \and
          J.~Bae
          \inst{\ref{inst9}}
          \and
          A.~Isella
          \inst{\ref{inst10}}
          \and
          N.~Kurtovic
          \inst{\ref{inst2}}
          \and
          L. M.~Pérez
          \inst{\ref{inst3}}
          \and
          P.~Pinilla
          \inst{\ref{inst7}}
          \and
          L.~Rampinelli
          \inst{\ref{inst4}}
          \and
          R.~Teague
          \inst{\ref{inst11}}
          \fnmsep
          }

   \institute{Université Côte d'Azur, Observatoire de la Côte d'Azur, CNRS, Laboratoire Lagrange\\
              \email{daniele.fasano@oca.eu}\label{inst1}
        \and
            Max-Planck-Institut für Astronomie, Königstuhl 17, 69117 Heidelberg, Germany\label{inst2}
        \and
            Departamento de Astronomía, Universidad de Chile, Camino El Observatorio 1515, Las Condes, Santiago, Chile\label{inst3}
        \and
            Dipartimento di Fisica, Università degli Studi di Milano, Via Celoria 16, Milano, Italy\label{inst4}
        \and
            National Astronomical Observatory of Japan, 2-21-1 Osawa, Mitaka, Tokyo 181-8588, Japan\label{inst5}
        \and
            Department of Astronomical Science, The Graduate University for Advanced Studies, SOKENDAI, 2-21-1 Osawa, Mitaka, Tokyo 181-8588, Japan\label{inst6}
        \and
            Mullard Space Science Laboratory, University College London, Holmbury St Mary, Dorking, Surrey RH5 6NT, UK\label{inst7}
        \and
            Center for Astrophysics, Harvard \& Smithsonian, 60 Garden Street, Cambridge, MA 02138, USA\label{inst8}
        \and
            Department of Astronomy, University of Florida, Gainesville, FL 32611, USA\label{inst9}
        \and
            Department of Physics and Astronomy, Rice University, 6100 Main Street, MS-108, Houston, TX 77005, USA\label{inst10}
        \and
            Department of Earth, Atmospheric, and Planetary Sciences, Massachusetts Institute of Technology, Cambridge, MA 02139, USA\label{inst11}
            }

   \date{Received April 1, 2025; accepted June 4, 2025}

  \abstract{The two giant protoplanets directly detected in the dust-depleted cavity of PDS 70 offer a unique opportunity to study the processes of ongoing planet formation. The planets have been detected both in IR thermal light and in the H$\alpha$ line, indicating that they are actively accreting material from their surroundings.}
   {We calibrate and analyse archival Atacama Large Millimetre/subMillimetre Array (ALMA) Band 6 and 7 observations of PDS~70 to detect circumplanetary material in independent datasets taken at different epochs in 2019, 2021 and 2023 and assess its possible motion.}
   {We perform 2D visibility modelling of the high resolution ALMA Band 6 ($\sim0.11\arcsec\times0.08\arcsec$) and Band 7 ($\sim0.05\arcsec\times0.05\arcsec$)  dust continuum emission of the outer disc. After subtracting the model from the data, we image the dust continuum emission in the cavity of PDS~70 at multiple epochs.}
   {We re-detect the compact dust emission around PDS 70 c in all our datasets in Band 6 and 7, with more than $3.8\sigma$ significance, and tentatively detect compact emission around PDS 70 b at $\sim3\sigma$ in the Band 6 datasets, with a peak emission of $59\pm17~\rm\mu Jy/beam$ and $46\pm14~\rm\mu Jy/beam$. We find the astrometric relative position of the compact emission around PDS 70 c to be consistent with the expected position of the planet in the 2019-2023 time range. We measure a peak flux difference of up to $64\pm34\rm~\mu Jy/beam$ at $1\sigma$ confidence level for the continuum emission coming from the region around PDS 70 c and perform a Bayesian test on our measurements, finding that they are not consistent with significant variable emission. We find no evidence of flux variability in the inner disc. We measure the dust mass of the material co-located with PDS 70 c and the inner disc to be in the range $0.008-0.063~M_\oplus$ and $0.04-0.31~M_\oplus$, respectively, consistent with previous measurements. Additionally, we obtain Band 6-7 spectral indices of $2.5\pm1.2$ and $3.2\pm0.5$ for the dust emission around PDS 70 c and in the inner disc, respectively, suggesting the inner disc emission is dominated by optically thin dust.} 
   {}

   \keywords{
                Planets and satellites: formation --
                Protoplanetary discs --
                Planet-disc interactions
               }
   \titlerunning{Inner disc and circumplanetary material in the PDS70 system}
   \maketitle

%
%-------------------------------------------------------------------

\section{Introduction}
\label{sec:Introduction}
%--------------------------------------------------------------------

In the past decade, observing facilities like the Very Large Telescope (VLT) and the Atacama Large Millimetre/submillimetre Array (ALMA) and detection techniques have pushed forward the field of planet formation by unveiling an astounding variety of substructures in protoplanetary discs, such as gaps, spirals, rings and clumps \citep{Garufi_2018, LongF_2018, Andrews_2020, Benisty_2023}. A possible interpretation for the origin of these substructures consists of newly born planets still forming in protoplanetary discs and interacting with them \citep{Bae_2023, Paardekooper_2023}. However, many of those substructures have only remained speculatively associated to planets, as the direct detection of planets has remained challenging with most candidate detections still debated \citep[e.g., ][]{Currie_2017, Rameau_2017}. So far only two protoplanets are firmly confirmed and are hosted in the cavity of the transition disc of PDS 70 \citep{Haffert_2019}. With a mass of $0.8~M_\sun$ and an age of $5.4~\rm Myr$ \citep{Muller_2018}, PDS 70 is a T Tauri star situated in the Upper Centaurus Lupus association \citep{Pecaut_Mamajek_2016}, at a distance of $112.4~\rm pc$ \citep{Gaia_2021}. The two giant protoplanets, PDS 70 b and c, have been detected through multiple tracers, confirming the planetary nature of their emission \citep{Keppler_2018, Muller_2018, Close_2025}, making PDS 70 an ideal system to directly study the interaction between the planets and the hosting disc \citep{Bae_2019, Toci_2020}. 

The disc surrounding PDS~70 shows a cavity, an outer ring with an inner shoulder and an over-brightness in the millimetre regime in the North-West \citep{Hashimoto_2012,Long_2018,Keppler_2019, Doi_2024}. Additionally, it features millimetre emission co-located with PDS 70 c, interpreted as a compact, circumplanetary disc (CPD), and diffuse dust emission detected around PDS 70 b \citep{Isella2019, Benisty_2021}. Based on the same observations, the system shows variability in the morphology of the inner disc \citep{Casassus_2022} as well as in the small dust reservoir of the inner disc, as shown by the spectral analysis of the James Webb Space Telescope (JWST) and \textit{Spitzer} data performed by \citet{Jang_2024, Perotti_2023, Liu_2025}. The inner disc emission is suggestive of dust filtration from the outer ring to the innermost regions of the system, with the largest grains trapped in the ring pressure maxima and the smaller grains able to filter through the gap and replenish the inner disc, as shown in the models from \citet{Pinilla_2024}.

In this paper, we present new high angular resolution ALMA Band 6 ($\sim0.15\times0.09\arcsec$) and 7 ($\sim0.07\times0.06\arcsec$) observations of PDS 70. With these new data, obtained at multiple epochs, we aim to re-detect and confirm the presence of the emission co-located with the planets and measure its relative astrometric motion compared to the expected planets' motion. In order to perform these measurements, we provide a detailed model of the outer disc, fitting for the geometrical and morphological properties of the disc in the visibility plane. Additionally, we analyse the continuum properties of the emission inside the cavity of PDS 70, providing estimates of the dust masses and the spectral index associated with the CPD around PDS 70 c and the inner disc.

This paper has the following structure: in Sec.~\ref{sec:Observations} we present the observations and describe the calibration procedure; in Sec.~\ref{sec:Methods} we introduce the methods we used to model the data, while in Sec.~\ref{sec:Results} we present our results and discuss them in Sec.~\ref{sec:Discussion}; we summarise our findings in Sec.~\ref{sec:Conclusions}.

\section{Observations}
\label{sec:Observations} 

\begin{table*}
\caption{Summary of the data used in this work. }             
\label{table:observation_summary} 
\centering                        
\footnotesize
\renewcommand{\arraystretch}{1.5}
\begin{tabular}{l c c c c c c r}  
\hline\hline                
Label & Project ID & Observation Date & Baselines & Frequency & Resolution & Max. Scale & References\\
 &    &  & [m] & [GHz]& [arcsec] & [arcsec] \\ 
\hline                        
B7 2019 & 2015.1.00888.S  & 2016 Aug 14-18  &  15–1462  &  344–355   & 0.16  & 3.23 &  \cite{Long_2018}\\
        & 2018.A.00030.S  & 2019 Jul 27-31  &  92–8547  &  346–355   &  0.03 & 0.53 &  \cite{Benisty_2021}\\
\hline                                   
B7 2021 & 2018.1.01774.S  & 2021 Jul 18-19  &  15-3696  &  330-346  & 0.07  & 1.4 &  This work \\
        & 2019.1.01138.S  & 2021 Aug 6  &  70-6855  &   344-357  & 0.03 & 0.52 &  This work\\
\hline                                   
B7 2023 & 2021.1.00782.S  & 2023 Mar 2  & 15-784   & 331-346  & 0.25 & 4.1 &  This work \\
        & 2022.1.01695.S  & 2023 Mar 2  &  15-784  &  343-357   & 0.28 & 3.65 &  \citet{Rampinelli_2025} \\
        & 2022.1.01695.S  & 2023 Jun 1-5  &  28-3638  &  343-357   & 0.06 & 0.996 &  \citet{Rampinelli_2025} \\
\hline
B6 2021 LF & 2019.1.01619.S  & 2020 Mar 6-16  &  46-364  &  217-233  & 0.40  & 6.21 & \cite{Facchini2021}\\
          & 2019.1.01619.S  & 2021 Jul 14-15 &  15.0-3396.4  & 217-233  & 0.1  & 1.8 &  \cite{Law2024}\\
\hline
B6 2021 HF & 2019.1.01619.S  & 2020 Mar 2-6  &  41-328  &  244-262 & 0.42  & 5.84 &  \cite{Facchini2021}\\
          & 2019.1.01619.S  & 2021 May 26-Jun 9  &  105-2400  & 244-262  & 0.10  & 3.25 &  \cite{Rampinelli_2024}\\
\hline     
\end{tabular}
\end{table*}

This work makes use of multiple datasets from different ALMA programs, that we categorized based on their observing frequency and the date of their long-baseline (LB) observations. Since our primary focus is to spatially resolve the compact emission around planet c and study its motion, we focus on the LB observations taken within one month. We do not apply the same restriction to the short-baseline (SB) observations as they do not spatially resolve the emission in the vicinity of the planets. A summary of all datasets used in this study is shown in Table~\ref{table:observation_summary}, while Table~\ref{table:observation_log} in the Appendix provides the observing log for the new ALMA datasets presented in this work for the first time. We performed the calibration using the version 6.2.1.7 of the CASA software \citep{CASA_2022}.

Combining all available Band 7 programs, we built three blocks of data that we label `B7 2019', `B7 2021' and `B7 2023'. B7 2019 uses LB data taken in 2019 and short baselines taken in 2016. B7 2021 combines SB and LB data taken in July and August 2021. Finally, B7 2023 includes SB and LB data taken in March and June 2023, respectively. In Band 6, we use two blocks of data, `B6 2021 LF' with a median frequency of 220 GHz with LB data taken in July 2021 and SB data taken in March 2020, and `B6 2021 HF' corresponding to a median frequency of 260 GHz, with LB data taken in June 2021 and SB data taken in March 2020. We detail the calibration procedure for all ALMA programs below and refer to Table~\ref{table:observation_summary} for the combination of programs that constitute the blocks of data described above. 

\underline{Band 7:} The calibration of data from 2015.1.00888.S and 2018.A.00030.S to build B7 2019 is described in \cite{Benisty_2021}, and for 2022.1.01695.S in \citet{Rampinelli_2025}. 
For data from 2018.1.01774.S; 2021.1.00782.S, we followed the 
iterative self-calibration following the exoALMA pipeline \citep{Loomis_2025}. First, spectral regions covering the following lines  
\ce{^13CO} ($3-2$, $\nu=330.59\,\mathrm{GHz}$),
\ce{CS} ($7-6$, $\nu=342.88\,\mathrm{GHz}$),
\ce{HC^15N} ($4-3$, $\nu=344.20\,\mathrm{GHz}$),
\ce{SO} ($8_8-7_7$, $\nu=344.31\,\mathrm{GHz}$),
\ce{H^13CN} ($4-3$, $\nu=345.34\,\mathrm{GHz}$), and
\ce{^12CO} ($3-2$, $\nu=345.80\,\mathrm{GHz}$),
were flagged and the remaining channels were averaged.
Then, an initial phase-only self-calibration was applied for each execution block (EB) and all EBs were aligned using the exoALMA alignment procedure \citep{Loomis_2025}, that regrids the EBs onto a common \textit{uv}-grid with natural weighting, retaining only overlapping grid cells and applying phase shifts to minimize visibility differences.
The EBs with $>4\%$ total flux deviation were rescaled to the total flux measured in a short-baseline EB with the highest signal-to-noise ratio.
The short-baseline EBs were concatenated and four rounds of phase-only self-calibration were executed with the solution intervals of EB-long, 360s, 120s, and 60s.
All EBs including the longer baselines were then combined.
Five rounds of phase-only self-calibration were similarly applied by shortening the solution intervals from EB-long to 30s.
Finally, after CLEANing down the data to the 1$\sigma$ threshold using the \texttt{tclean()} task of the CASA software \citep{Hogbom_1974, Cornwell_2008}, amplitude and phase self-calibration were performed with an EB-long solution interval.
As shown in Table~\ref{table:observation_summary}, the 2018.1.01774.S and 2021.1.00782 data were then combined with other dataset to form the B7 2021 and B7 2023 data blocks.

We proceeded similarly for program 2019.1.01138.S (single EB). As a first step, spectral lines in the Band 7 frequency range were flagged \citep{Rampinelli_2025}: \ce{HC^15N} ($4-3$, $\nu=344.20\,\mathrm{GHz}$)
\ce{SO} ($8_8-7_7$, $\nu=344.31\,\mathrm{GHz}$),
\ce{H^13CN} ($4-3$, $\nu=345.34\,\mathrm{GHz}$),
\ce{HCN} ($4-3$, $\nu=354.51\,\mathrm{GHz}$),
\ce{^12CO} ($3-2$, $\nu=345.80\,\mathrm{GHz}$),
\ce{HCO^+} ($4-3$, $\nu=356.73\,\mathrm{GHz}$).
The remaining unflagged data were then averaged into 250 MHz-wide channels. An initial phase-only self-calibration round over the entire EB duration was applied to the EB, combining all scans and spectral windows. 

To build the B7 2021 block, the July 2021 EB from 2018.1.01774.S was then aligned to the 2019.1.01138.S EB using the exoALMA alignment procedure \citep{Loomis_2025}. No flux rescaling was applied, as total flux differences remained within 4\%, accounting for the slight frequency offset between the two programs (mean frequency of 338 GHz for 2018.1.01774.S and 350 GHz for 2019.1.01138.S). The two EBs were then concatenated, and phase-only self-calibration was applied in multiple rounds, progressively shortening the solution intervals (EB-long, 360s, 120s, 60s, 30s, 18s). After this, the data were cleaned down to $1\sigma$, and two rounds of amplitude and phase self-calibration were performed, combining polarizations and spectral windows. The first round used EB-long solution intervals, while the second used scan-long intervals.

To build B7 2023, the SB datasets from program 2021.1.00782.S were combined to the LB data from program 2022.1.01695.S following a similar procedure for the alignment as described above. No flux rescaling applied, as total flux differences remained within 4\% accounting for the slight frequency offset (mean frequency of 338 GHz for 2021.1.00782.S and 350 GHz for 2022.1.01695.S).

\underline{Band 6:} For the Band 6 datasets, the calibration procedures are detailed in \cite{Facchini2021}, \cite{Law2024}, and \cite{Rampinelli_2024}. These datasets include two distinct spectral setups: a lower-frequency one at $\sim220$\,GHz (B6 2021 LF) and a higher-frequency setup centred at $\sim260$\,GHz (B6 2021 HF).

Finally, all datasets were averaged to one channel per spectral window and binned into 30s time intervals. They were then all aligned using the exoALMA alignment procedure \citep{Loomis_2025}, with the B7 2023 dataset serving as the reference. 

\section{Methods}
\label{sec:Methods} 

As the emission of the cavity is faint and partly overlapping with the outer disc, we first model the outer disc. After subtracting the model from the data, we obtain residual maps that contain a clear signal from the cavity. We characterize the morphological structure of the continuum emission with the code \texttt{galario} \citep{Tazzari_2018}, fitting the visibility data assuming a 2D parametric intensity model and following a Markov Chain Monte Carlo (MCMC) approach implemented with the package \texttt{emcee} \citep{Foreman-Mackey_2013}. Following this procedure, we can constrain both the geometrical parameters of the outer disc, specifically the inclination $i$, the position angle PA and the offsets $(\Delta\rm RA, \Delta Dec)$ between the disc and the phase centres, as well as the morphological parameters specific to the chosen model.

    \begin{table*}
    \caption{Disc continuum properties and geometrical parameters.}             
    \label{table:Disc continuum properties}      
    \centering        
    \footnotesize
    \renewcommand{\arraystretch}{1.5}
    \begin{tabular}{c c c c c c c c c c}        
    \hline\hline           
    Dataset & Robust, Beam & PA$_{\rm beam}$ & Peak Flux & Flux Density & RMS noise & $i$ & PA & $\Delta\rm RA$ & $\Delta\rm Dec$\\
     & ,[mas$\times$mas] & [deg] & [mJy beam$^{-1}$] & [mJy] & [$\mu$Jy beam$^{-1}$] & [deg] & [deg] & [mas] & [mas]\\
    \hline                       
       B7 2019 & 1.0, 52 $\times$ 45 & 69 & $0.96\pm0.10$ & $176\pm18$ & 15 & $49.94_{-0.01}^{+0.02}$ & $160.46_{-0.01}^{+0.06}$ & $-19.2_{-0.2}^{+0.2}$ & $2.3_{-0.2}^{+0.2}$\\     
       B7 2021 & 1.0, 60 $\times$ 45 & 79 & $1.18\pm0.12$ & $187\pm19$ & 25 & $49.83_{-0.02}^{+0.02}$ & $160.00_{-0.04}^{+0.05}$ & $-19.4_{-0.2}^{+0.2}$ & $2.4_{-0.2}^{+0.2}$\\
       B7 2023 & 0.0, 72 $\times$ 62 & $-90$ & $1.85\pm0.19$ & $187\pm19$ & 19 & $49.82_{-0.01}^{+0.01}$ & $160.40_{-0.02}^{+0.01}$ & $-18.6_{-0.1}^{+0.1}$ & $0.8_{-0.9}^{+0.9}$\\
       B6 2021 LF & 0.0, 115 $\times$ 89 & $-84$ & $1.46\pm0.15$ & $58\pm6$ & 16 & $49.98_{-0.01}^{+0.07}$ & $160.94_{-0.02}^{+0.04}$ & $-5.5_{-0.1}^{+0.3}$ & $2.4_{-0.1}^{+0.3}$\\
       B6 2021 HF & 0.0, 110 $\times$ 79 & $-86$ & $1.48\pm0.15$ & $68\pm7$ & 13 & $50.097_{-0.005}^{+0.021}$ & $160.64_{-0.01}^{+0.04}$ & $-14.4_{-0.2}^{+0.2}$ & $2.2_{-0.2}^{+0.2}$\\
    \hline                                   
    \end{tabular}
    \end{table*}

We define our model brightness distribution as follows:

\begin{equation}
    I(r,\theta) = \sum_{i=1}^{2} GR(f_i, r_i, \sigma_{r,i}) + GA(f^{\rm a}_1, r^{\rm a}_1, \sigma^{\rm a}_{r,1}, \theta^{\rm a}_1, \sigma^{\rm a}_{\theta,1})
\end{equation}
where $GR(f, r, \sigma)$ is an axisymmetric Gaussian ring centred in $r$ with radial width $\sigma$ and peak flux $f$, while $GA(f, r, \sigma_r, \theta, \sigma_\theta)$ is a Gaussian arc centred in $(r,\theta)$ with radial width $\sigma_r$, azimuthal width $\sigma_\theta$ and peak flux $f$. We consider two Gaussian components to model the outer ring in order to replicate the inner shoulder seen in the observations, while we use the arc to reproduce the overdensity in the North West of the ring. 

We perform a preliminary exploration of the parameter space with \texttt{emcee} using uniform priors, sampling all the parameters linearly with the exception of $f_1$, $f_2$ and $f_1^a$, which are sampled logarithmically. As initial step we use 100 walkers over $\sim10^3$ steps. We then set the best fit parameters, computed as the median of the posterior distribution from this preliminary exploration, as initial guesses for the fiducial \texttt{emcee} run. For each dataset we set up 250 walkers and $10^4$ steps. We discard the first $25\%$ steps as burn-in. We show the full set of best fit parameters in Table~\ref{table:galario results} and report the adopted priors in Appendix~\ref{App: Visibility Modelling}. We then subtract the best fit \texttt{galario} model evaluated at the same $uv$ points of the observations from the observed data in order to obtain the residuals. We assume that the uncertainties introduced with this subtraction are dominated by the ALMA flux uncertainties we discuss in Sec.~\ref{sec:Results}. Additionally, we perform a test run with 100 walkers and 1000 steps on the B7 2019 dataset adding a central Gaussian $G(f_0, \sigma_{r,0})$ to our model. We check that the resulting bestfit parameters are within $1\sigma$ uncertainties of those obtained in our fiducial run, which implies that the errors introduced by not fitting for the inner disc in our models are negligible.

We image the data, the best fit \texttt{galario} model and the residuals for each dataset using the \texttt{tclean()} task of the CASA software using a robust parameter of 1.0 (B7 2019, B7 2021) and 0.0 (B7 2023, B6 2021 LF, B6 2021 HF), chosen to provide the best balance between resolution and sensitivity, and produce $1024\times1024$ pixels images with a pixel size of $0.01\arcsec$ in Band 7 and $512\times512$ pixels images with a pixel size of $0.02\arcsec$ in Band 6. To avoid negative point-source components in the CLEAN models, we create tailored masks for each image around the $3\sigma$ contours of the outer ring, inner disc and the compact emission co-located with PDS 70 c and CLEAN the images down to $1\sigma$ using a very conservative \texttt{gain} parameter of $0.02$. Our fiducial images have the following resolutions (in mas): B7 2019, 52x45; B7 2021, 60x45; B7 2023, 72x62; B6 2021 LF, 115x89; B6 2021 HF, 110x79, see Tab.\ref{table:Disc continuum properties}. 

In the following, when measuring the peak intensity we estimate its uncertainty by adding in quadrature the RMS noise of the image together with the $10\%$ ALMA flux calibration error for Band 6 and 7\footnote{see Sect. 10.2.6 in the ALMA Technical Handbook https:/almascience.nrao.edu/proposing/technical-handbook/}. When measuring flux densities in a given region, instead, we measure the flux density in the same area centred on random locations outside of the disc emission and compute the standard deviation, according to the procedure presented in \cite{Rampinelli_2024} and add it in quadrature to the $10\%$ ALMA flux calibration error.

\section{Results}
\label{sec:Results}

    \begin{figure*}[!h]
    \centering
    \includegraphics[width=0.9\textwidth]{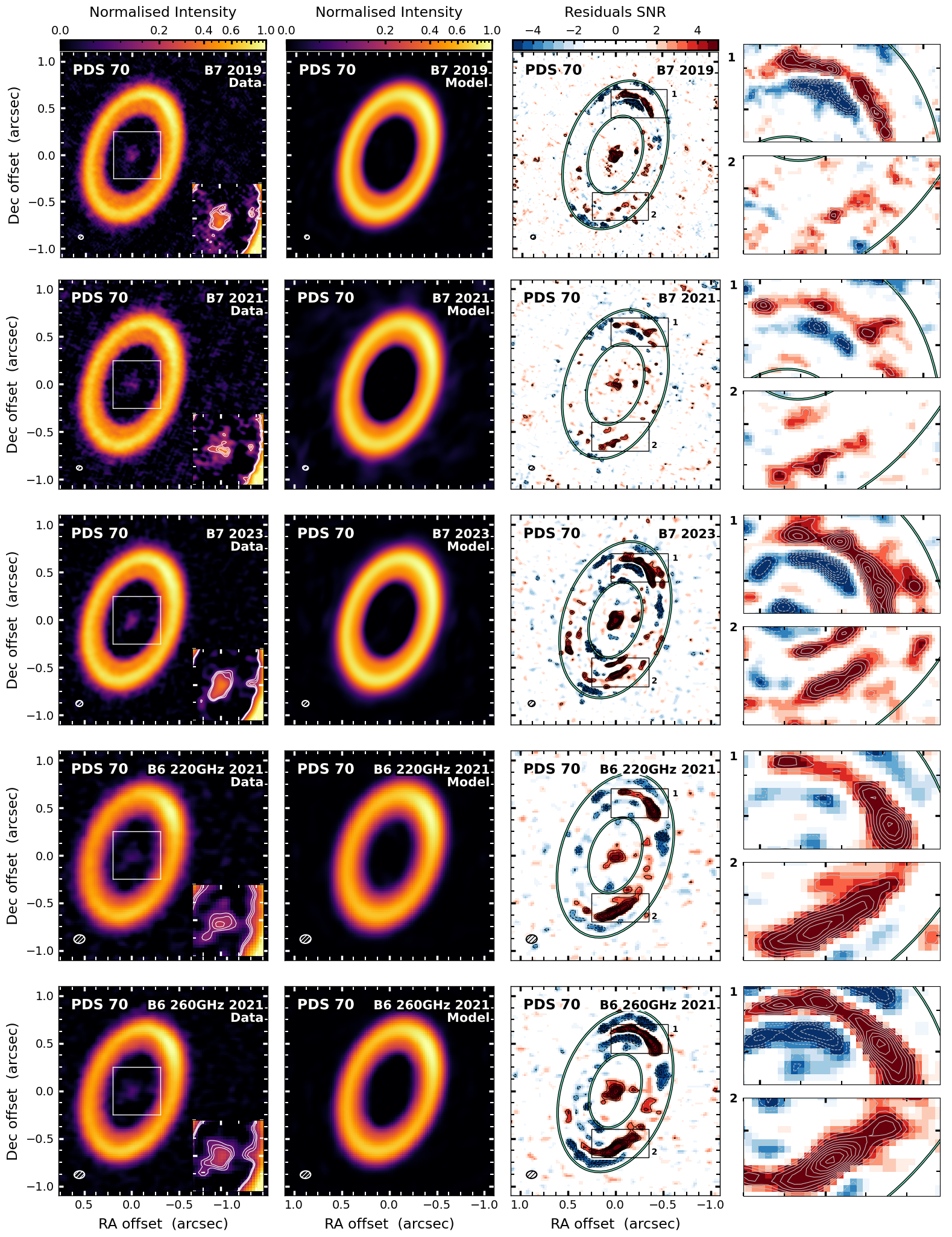}
 \label{fig:Summary plots}
 \caption{Left to right: Continuum images with insets showing a zoom in the cavity; CLEANed \texttt{galario} model images; CLEANed residual images; zoom on the North (1) and South (2) sides of the outer ring residuals. Contours start at $3\sigma$ and increase by $1\sigma$. The green ellipses approximate the $7\sigma$ contours of the continuum data. The datasets are ordered from Band 7 to Band 6 and by epoch from top to bottom.}
 \label{fig:Summary plots}
    \end{figure*}

We show in Fig.~\ref{fig:Summary plots} the main results of the continuum analysis. In the left column we present the continuum images for all datasets. We report the main continuum properties together with the disc geometrical parameters in Table~\ref{table:Disc continuum properties}. The values we find for the inclination and the position angle are within $1\degree$ of those obtained in \citet{Keppler_2019, Benisty_2021, Aizawa_2024}, with this difference most likely due to the different methodologies used. \citet{Benisty_2021} also estimate an offset of $\Delta\rm RA = 12 mas$ and $\Delta\rm Dec = 15 mas$ using \texttt{frank} \citep{Jennings_2020}, similar to the values \citet{Aizawa_2024} obtain with their methodology. Both methods assume the disc emission to be axisymmetric, which might explain the discrepancy with the offsets ($\Delta\rm RA, \Delta\rm Dec$) we obtain from our 2D fit.  In all our datasets we observe the well-known structure of PDS~70: a cavity, an outer ring with an inner shoulder and an asymmetric overbrightness in the North-West. However, the Band 6 datasets show the presence of a fainter overbrightness also in the South-West of the outer ring. While the morphological structure of the ring appears unchanged at different epochs, the emission in the cavity, seen in an inset for each plot, shows a variable morphology. This can be in part explained by the Signal to Noise Ratio (SNR) inside the cavity, defined as the ratio of the inner disc peak emission and the image RMS, for the 2021 dataset, which is $\sim35\%$ and $\sim50\%$ lower than the SNR of the 2019 and 2023 datasets, respectively, and the different beam size in the 2023 dataset. In the middle and right column, we show the CLEAN images of the \texttt{galario} model and residuals. The residual images should be interpreted differently in two distinct spatial regions, the outer ring and the cavity that are indicated using the $7\sigma$ contours of the continuum emission from the observations. The residuals in the outer disc show deviations from our disc model, while the residuals in the cavity provide information on the emission from the inner disc and the dust in the vicinity of the two planets. 

\paragraph{\textit{Outer ring}} The best fit models for the outer ring feature a broad ring at $r_1\sim0.64-0.68\arcsec$ ($72-77\rm~au$) with a radial half-width of $\sigma_1\sim0.07-0.09\arcsec$ ($8-10\rm~au$), a narrow ring centred in $r_2\sim0.47-0.50\arcsec$ ($53-56\rm~au$) with a radial half-width of $\sigma_2\sim0.03-0.06\arcsec$ ($3-7\rm~au$), representing the shoulder, and an arc at $r_1^{\rm a}\sim0.63-0.70\arcsec$ ($71-79\rm~au$) and $\theta_1^{\rm a}\sim135-142\degree$, which accounts for the asymmetric emission (see Table.~\ref{table:galario results}). Here we note that an azimuthal shift of $7\deg$ from the 2019 to the 2023 epoch matches the Keplerian speed of PDS 70 c, suggesting that this substructure might be related to the planet. 

Most of the residuals in the outer ring of the B7 2019 and 2021 datasets are lower than $3\sigma$, showing that our model is able to explain the global structure of the outer disc of PDS 70. The B7 2023 and both B6 2021 datasets, instead, show stronger residuals along the major axis. Our results are consistent with the analysis of PDS 70 presented in \citet{Aizawa_2024}, where they first apply an axisymmetric model to the emission of the disc and then fit a parametric model to the asymmetry in the residuals, finding a similar pattern of positive and negative residuals in the region surrounding the asymmetry. We note in all our datasets the presence of strong ($>5\sigma$) residuals co-located with the asymmetry in the North-West. These residuals are characterised by multiple positive and negative peaks up to $\sim6\sigma$ (B7 2021), $\sim7\sigma$ (B7 2019), $\sim9\sigma$ (B7 2023, B6 2021 LF) and $\sim11\sigma$ (B6 2021 HF). This implies that the morphology of the asymmetry is more complex than a simple 2D Gaussian in radial and azimuthal direction. After convolving the three Band 7 images with a Gaussian beam to the resolution of the B7 2023 dataset using the \texttt{imsmooth} task of the CASA software, we check that the morphology of the residuals is consistent across the different epochs on a qualitative level, with the main differences most likely due to the different \textit{uv} coverage of the datasets. Additionally, both B6 datasets show similarly strong ($\sim10\sigma$) localised residuals in the South-West of the outer ring. The presence of these residuals localised only around the semi-major axis can be possibly explained as optical depth effect arising in inclined discs \citep{Doi_2021}.

    \begin{figure*}
    \centering
    \resizebox{\hsize}{!}{
        \begin{subfigure}[t]{\textwidth}
        \centering
        \includegraphics[width=\textwidth]{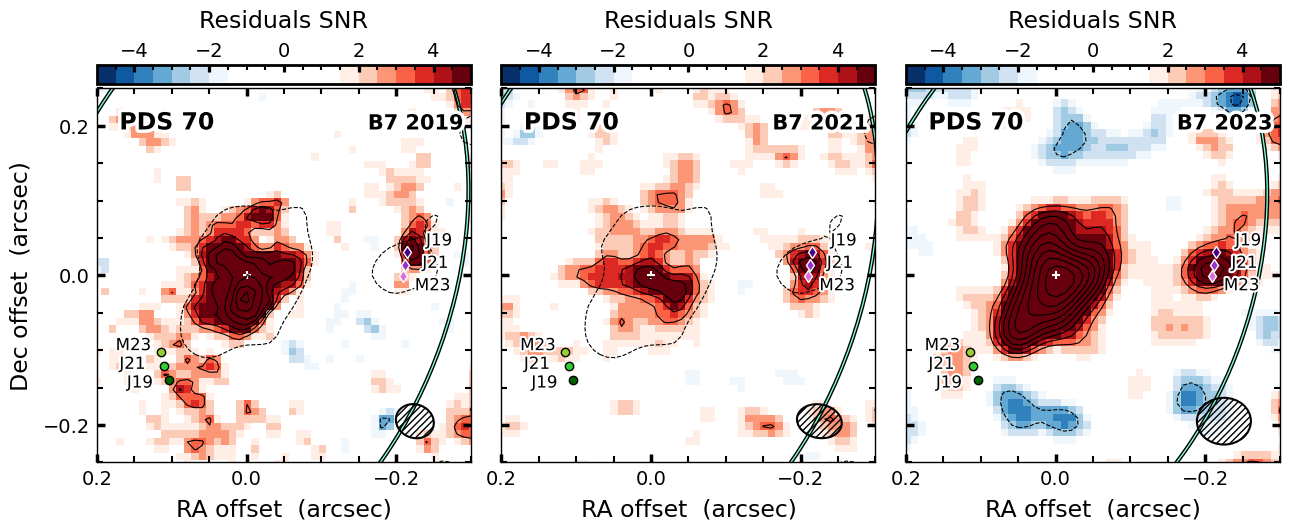}
        \caption{
              }
         \label{fig:Residual cavity all B7}
        \end{subfigure}
        }
     \resizebox{\hsize}{!}{
        \hfill 
        \begin{subfigure}[t]{\textwidth}
        \centering
        \includegraphics[width=0.7\textwidth]{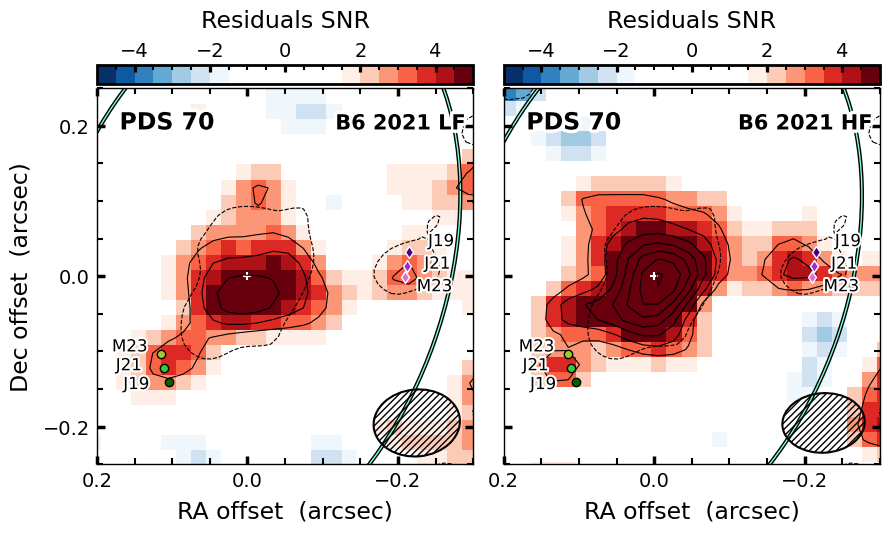}
        \caption{
              }
         \label{fig:Residual cavity all B6}
         \end{subfigure}
        }
    \caption{Zoom of the residuals in the cavity from Fig.\ref{fig:Summary plots}. a) Band 7 2019 (left panel), 2021 (middle panel) and 2023 (right panel) datasets. b) Band 6 2021 datasets at $\rm220~GHz$ (left panel) and $\rm260~GHz$ (right panel). The solid contours start from the $3\sigma$ emission and increase by $1\sigma$, while the  dashed contours correspond to the $3\sigma$ emission from the B7 2023 dataset and are overlaid to the other datasets for comparison. The white plus sign marks the centre of the disc, corrected with the \texttt{galario} offset. The green circles and purple diamonds represent the expected position in July 2019 (dark), July 2021 (medium) and May 2023 (light) obtained from the orbital fit of PDS 70 b and c \citep{Wang_2021}, respectively. The error bars are smaller than the markers. 
              }
    \label{fig:Residual cavity all}
   \end{figure*}

    \begin{table*}
    \caption{$c_{\rm smm}$ continuum properties}
    \label{table:CPD continuum properties}      
    \centering         
    \footnotesize
    \renewcommand{\arraystretch}{1.5}
    \begin{tabular}{c c c c c c} 
    \hline\hline    
    Dataset & Peak Flux & SNR & $F_\nu$ & $M_{\rm d} (1~\rm mm)$ & $M_{\rm d} (1~\rm \mu m)$\\
     & [$\mu$Jy beam$^{-1}$] & & [$\mu$Jy] & [$M_{\oplus}$] & [$M_{\oplus}$]\\
    \hline
       B7 2019 & $94\pm18$ & 6.4 & $95\pm26$ & 0.008 & 0.035\\     
       B7 2021 & $139\pm28$ & 5.6 & $168\pm39$ & 0.014 & 0.063\\
       B7 2023 & $127\pm23$ & 6.7 & $137\pm23$ & 0.011 & 0.052\\
       B6 2021 LF & $54\pm17$ & 3.8 & $52\pm15$ & 0.009 & 0.042\\
       B6 2021 HF & $54\pm14$ & 4.3 & $66\pm13$ & 0.009 & 0.039\\
    \hline                                 
    \end{tabular}
    \end{table*}

    \paragraph{\textit{Circumplanetary material}} In Fig. \ref{fig:Residual cavity all}  we show the residuals inside the cavity. We detect compact dust emission co-located with PDS 70 c ($c_{\rm smm}$ for brevity) in all our datasets with at least a $~5\sigma$ significance for Band 7 and a $~3.8\sigma$ significance for Band 6. As previous publications were based on the same datasets, this allows to re-detect with independent and new datasets the presence of dust in the vicinity of the planets, confirming previous observations \citep{Isella2019, Benisty_2021, Casassus_2021}. The compact source $c_{\rm smm}$ is unresolved in all the datasets at the resolution of our observations, consistent with being originated from a compact object. We report the continuum properties of $c_{\rm smm}$ in Table~\ref{table:CPD continuum properties}. For the B7 2019 dataset we measure a peak flux emission of $94\pm18~\rm\mu Jy/beam$, compatible within $1\sigma$ with the value of $81\pm5~\rm\mu Jy/beam$ obtained for the dataset LB19$+$SB16 by \citet{Benisty_2021}, where our larger uncertainty takes into account the ALMA flux calibration errors. We then quantify the variability of $c_{\rm smm}$ across the different epochs in Band 7, using the peak intensity as the emission is unresolved. In order to perform a consistent comparison between the different datasets, we first use the task \texttt{imsmooth} of the CASA software to convolve the images with a Gaussian beam to the resolution of the B7 2023 images. We measure a peak intensity for $c_{\rm smm}$ of $88\pm17~\rm\mu Jy/beam$, $153\pm29~\rm\mu Jy/beam$ and $127\pm23~\rm\mu Jy/beam$ for B7 2019, B7 2021 and B7 2023, respectively, within $1\sigma$ error bars of the estimates from our fiducial images. Thus, we measure a tentative ($\sim2\sigma$) peak intensity variability of $64\pm34\rm~\mu Jy/beam$ between the 2019 and 2021 epochs, and a peak intensity difference of $-26\pm37\rm~\mu Jy/beam$ and $39\pm29\rm~\mu Jy/beam$ between the 2021,2023 and 2019, 2023 epochs, respectively. However, as these values are taken at $1\sigma$ confidence level, they can still be compatible with no variable emission. We additionally perform a Bayesian test using \texttt{emcee}, assuming that the peak intensity of $c_{\rm smm}$ is a constant value $F_0$ across the three epochs and considering the presence of an excess noise $\sigma_0$, so that the uncertainty of each peak flux measurement is given by $\sigma = \sqrt(\sigma_{\rm obs}^2+\sigma_0^2)$, with $\sigma_{i,\rm obs}$ the uncertainty of the measured peak flux. If our measurements are compatible with a constant flux, we expect to find $\sigma_0=0$, whether a non zero $\sigma_0$ would indicate the presence of variability. We setup \texttt{emcee} with 100 steps and let it run for 4000 steps, sampling $F_0$ linearly and $\sigma_0$ logarithmically, with priors of $[90, 200]~\rm\mu Jy/beam$ and $\rm[log_{10}1,log_{10}1000]$, respectively. We find $1\sigma$ confidence intervals of $[101,133]~\rm\mu Jy/beam$ and $[2,35]$ for $F_0$ and $\sigma_0$, respectively, further suggesting that the emission from $c_{\rm smm}$ is not variable.

Assuming that the emission is optically thin, the dust mass can be estimated following \citet{Hildebrand_1983}:

\begin{equation}
\label{Mass}
    M_{\rm dust} = \frac{d^2F_\nu}{\kappa_\nu B_\nu(T)}
\end{equation}
where $d$ is the distance to the source, $\nu$ the observed frequency,
$F_\nu$ the integrated flux density, $\kappa_\nu$ the dust absorption coefficient, $B_\nu$ the Planck function and $T$ the temperature of the source. If the emission is optically thick, the mass estimate from Eq.\eqref{Mass} should be considered only as a lower limit. We compute the integrated flux density $F_\nu$ inside an elliptical area with twice the beam semi-major and semi-minor axes centred on the brightest pixel. We follow the same approach of \cite{Benisty_2021} to estimate the dust mass of $c_{\rm smm}$ for all of the epochs considered in this work, assuming that the emission is due to either $1~\rm mm$ or, mimicking the case where the larger grains are trapped in the ring and only the smaller grains coupled with the gas can flow inside the cavity \citep{Bae_2019}, $1~\rm \mu m$ dust sized grains. In the former case we assume a typical dust opacity of $3.63~\rm cm^2g^{-1}$ \citep[DSHARP composition,][]{Birnstiel_2018}. Otherwise, we consider a dust opacity of $0.79~\rm cm^2g^{-1}$ \citep[DSHARP composition,][]{Birnstiel_2018}, typically associated with $1~\rm \mu m$ particles. We assume a temperature of $T=26~\rm K$, considering the contribution of viscous heating and irradiation from the accreting star and planet \citep[for a detailed derivation see][]{Benisty_2021} and find dust masses consistent with those obtained by \citet{Benisty_2021} (Table~\ref{table:CPD continuum properties}).

Around PDS 70 b we detect diffuse emission ($\sim3\sigma$) in the Band 7 datasets, as in  \cite{Benisty_2021}. In the Band 6 datasets, instead, we marginally detect ($\sim3\sigma$) for the first time compact emission co-located with the predicted position of PDS 70 b, connected to the inner disc, with a peak emission of $59\pm17~\rm\mu Jy/beam$ and $46\pm14~\rm\mu Jy/beam$ for B6 2021 LF and B6 2021 HF, respectively. The non-detection of a compact source in Band 7 might be associated to a low local spectral index. We provide an upper limit to this spectral index by measuring the peak emission inside the $3\sigma$ contours of the compact emission in B6 2021 LF and B6 2021 HF, and in the same region of B7 2021 after smoothing the image with a Gaussian beam to the same resolution of B6 2021 LF using the task \texttt{imsmooth} of the CASA software. We measure a peak intensity of $106\pm40\rm~\mu Jy/beam$ in Band 7, and compute a conservative $3\sigma$ spectral index upper limit\footnote{In this estimate we use the face value for the Band 6 flux and the face value plus $3\sigma$ for the non detection in Band 7.} of $\alpha<3.0$ and $\alpha<5.8$ for the lower and higher frequency Band 6 datasets, respectively. However, we note that these upper limits should be treated with care as the compact source is not well separated from the inner disc in Band 6 and the emission could be contaminated by material from the inner disc.

    \begin{table*}
    \caption{Inner disc continuum properties}             
    \label{table:Inner disc continuum properties}    
    \centering         
    \footnotesize
    \renewcommand{\arraystretch}{1.5}
    \begin{tabular}{c c c c c c} 
    \hline\hline           
    Dataset & Peak Flux & SNR & $F_\nu$ & $M_{\rm d} (1~\rm mm)$ & $M_{\rm d} (1~\rm \mu m)$\\
     & [$\mu$Jy beam$^{-1}$] & & [$\mu$Jy] & [$M_{\oplus}$] & [$M_{\oplus}$]\\ 
    \hline 
       B7 2019 & $138\pm20$ & 9.3 & $767\pm109$ & 0.077 & 0.353\\    
       B7 2021 & $149\pm29$ & 6.0 & $859\pm135$ & 0.089 & 0.407\\
       B7 2023 & $225\pm29$ & 11.8 & $785\pm84$ & 0.081 & 0.372\\
       B6 2021 LF & $95\pm19$ & 5.8 & $210\pm25$ & 0.046 & 0.211\\
       B6 2021 HF & $115\pm17$ & 9.1 & $285\pm32$ & 0.046 & 0.211\\
    \hline    
    \end{tabular}
    \end{table*}

\paragraph{\textit{Inner disc}} We perform the same continuum analysis on the inner disc and report our results in Table~\ref{table:Inner disc continuum properties}. In this case the emission is resolved, so we create a single elliptical mask covering the $3\sigma$ emission of the inner disc in B7 2023 and apply it to all the Band 7 datasets. We repeat the process for Band 6 using B6 2021 HF as the reference dataset for the mask. The flux density we obtain for the B7 2019 dataset is within the error bars of the value measured in \citet{Benisty_2021}. 
We find no evidence of variability for the inner disc of PDS 70, as our flux density estimates are consistent within error bars across the different epochs we consider. We measure the dust mass of the inner disc following the same procedure we used for $c_{\rm smm}$, considering an average temperature of $T=22\rm~K$ from the model used in \citet{Benisty_2021}. Our estimates are consistent with the range $0.08-0.36~M_\oplus$ reported in \citet{Benisty_2021}, with the dust masses extracted from the Band 6 observations being roughly half of those from Band 7. In Fig.~\ref{fig:Residual cavity all} we visually compare the morphology of the inner disc by overlaying the $3\sigma$ contour of the B7 2023 residuals to all the other datasets. If we approximate the $1\sigma$ mask with a Gaussian, the radius of the inner disc is in the range $0.10-0.14~\rm arcsec$ ($11-16\rm~au$), with the B7 2021 datasets likely underestimating the extent of the inner disc due to the lower SNR and the Band 6 featuring a slightly larger extent compared to the other datasets, probably due to the lower angular resolution. \citet{Casassus_2022} report two clumps in the inner disc at $0.03-0.04~\rm arcsec$ (Clump 1) and $0.08~\rm arcsec$ (Clump 2) at two different epochs (2017 and 2019). We re-detect Clump 1 at $\sim7\sigma$ and Clump 2 at $\sim4\sigma$ in the B7 2019 dataset (see left panel of Fig.~\ref{fig:Inner disc clumps}). However, we note that subtracting the test model that includes the central Gaussian we do not detect any clump with at least a $3\sigma$ significance in the region of the inner disc, similarly to the residuals in \citet{Aizawa_2024}. Due to the lower SNR of the B7 2021 dataset, a direct comparison with the previous epoch is not straightforward, making it unclear if the morphology of the inner disc is due to noise fluctuations or dust substructures. In the B7 2021 dataset, we do not detect clumps in the regions where we expect to see Clump 1 and Clump 2 assuming Keplerian rotation \citet{Casassus_2022} (see Appendix~\ref{App: Inner Disc Clumps}). We do not resolve any clump in the B7 2023 datasets, likely due to the lower angular resolution.

\section{Discussion}
\label{sec:Discussion}

\subsection{$c_{\rm smm}$ astrometry}

    \begin{table*}
    \caption{$c_{\rm smm}$ astrometry}          
    \label{table:CPD astrometry}
    \centering          
    \footnotesize
    \renewcommand{\arraystretch}{1.5}
    \begin{tabular}{c c c | c c | c c}       
    \hline\hline      
    & \multicolumn{2}{c}{{\bf Predicted}} & \multicolumn{2}{c}{{\bf Measured}} & \multicolumn{2}{c}{{\bf Displacement}} \\
    Dataset & $\Delta \rm RA$ & $\Delta \rm Dec$ & $\Delta \rm RA$ & $\Delta \rm Dec$ & $\Delta \rm RA$ & $\Delta \rm Dec$\\
     & [mas] & [mas] & [mas] & [mas] & [mas] & [mas]\\ 
    \hline                      
       2019 & $-214.8\pm0.3$ & $31.8\pm0.4$ & $-227\pm9$ & $59\pm9$ & $0$ & $0$\\
       2021 & $-212.1\pm0.6$ & $14.1\pm0.7$ & $-217\pm12$ & $25\pm12$ & $10\pm15$ & $-34\pm15$\\
       2023 & $-209.7\pm1.1$ & $-1.3\pm1.0$ & $-215\pm12$ & $17\pm12$ & $12\pm15$ & $-42\pm15$\\
    \hline     
    \end{tabular}
    \tablefoot{The offset $\Delta$RA is computed as $\Delta\rm RA\cos{(\rm DEC)}$, accounting for the World Coordinate System (WCS) projection. In the last two columns we show the components of the relative astrometric displacement with respect to the measured position of the 2019 dataset.}
    \end{table*}

We can use ALMA interferometric data to compute the position of $c_{\rm smm}$ and a possible motion with time. We focus only on the Band 7 datasets for this analysis due to their higher resolution. Although we cannot observe the star nor the planets themselves, preventing us to obtain absolute astrometric measurements, we can still define a relative coordinate system. We decide to centre our reference frame on the centre of the disc corrected with the \texttt{galario} offset. By doing so we assume that the centre fitted with \texttt{galario} coincides with the location of the star. We compute the expected position of both PDS 70 b and c using the orbital fitting tool\footnote{https://whereistheplanet.com} presented in \citet{Wang_2021} and overlay them in Fig.~\ref{fig:Residual cavity all}. The emission from $c_{\rm smm}$ overlaps with the expected planet positions at all three epochs. 

To quantify the possible motion of $c_{\rm smm}$, we fit a Gaussian to the CLEANed residual images in the region around $c_{\rm smm}$ using the CASA task \texttt{imfit} and use the coordinates of the centre resulting from the fit to estimate its position at different epochs. In Table~\ref{table:CPD astrometry} we report the offset in RA and Dec of both the expected position of PDS 70 c and the measured position of $c_{\rm smm}$. The astrometric accuracy of the relative position of $c_{\rm smm}$ between epochs is directly related to the accuracy of the relative alignment between the different datasets. For the angular resolution of our observations, we estimate this accuracy to be within $1~$mas, based on the offset uncertainties from our \texttt{galario} best fit models (see Table~\ref{table:galario results}. To this systematic uncertainty, we add in quadrature the uncertainty related to the identification of the peak of PDS 70 c, which can be computed as $\theta\rm _{FWHM}/SNR/0.9$, with the assumption that the emission is close to a point-source, where we consider the Full Width Half Maximum (FWHM) of the beam ($\theta\rm _{FWHM}$), the SNR of $c_{\rm smm}$ and a 0.9 factor to account for a $10\%$ signal decorrelation \footnote{see Sect. 10.5.2 in the ALMA Technical Handbook https:/almascience.nrao.edu/proposing/technical-handbook/}. We choose to employ a conservative estimate of this accuracy by considering the SNR of $c_{\rm smm}$ instead of that of the entire image. With a SNR of $6.4$, $5.6$ and $6.7$ and a beam FWHM of $52$, $60$ and $72~\rm mas$ for the 2019, 2021 and 2023 datasets, respectively, we find an astrometric accuracy of $9~\rm mas$, $12~\rm mas$ and $12~\rm mas$, respectively. We obtain a total displacement of $35\pm15~\rm mas$ ($\sim2\sigma$ significance) between the 2019 and 2021 epochs and $44\pm15~\rm mas$ ($\sim3\sigma$ significance) between the 2019 and 2023 epochs, with the peak intensity moving compatibly to the orbital solution of planet c by \citet{Wang_2021}. We note that \citet{Close_2025} report H$\alpha$ detections of the two planets of PDS 70 with one epoch (2023) overlapping with those considered in this work and consistent with the location of $c_{\rm smm}$. This suggests that $c_{\rm smm}$ is possibly co-moving with the planet, further strengthening the circumplanetary material scenario suggested in \citet{Benisty_2021} but further observations at high angular resolution in a few years will be necessary to confirm such tentative motion.

\subsection{$c_{\rm smm}$ variability}

\begin{figure}
   \centering
   \includegraphics[width=\hsize]{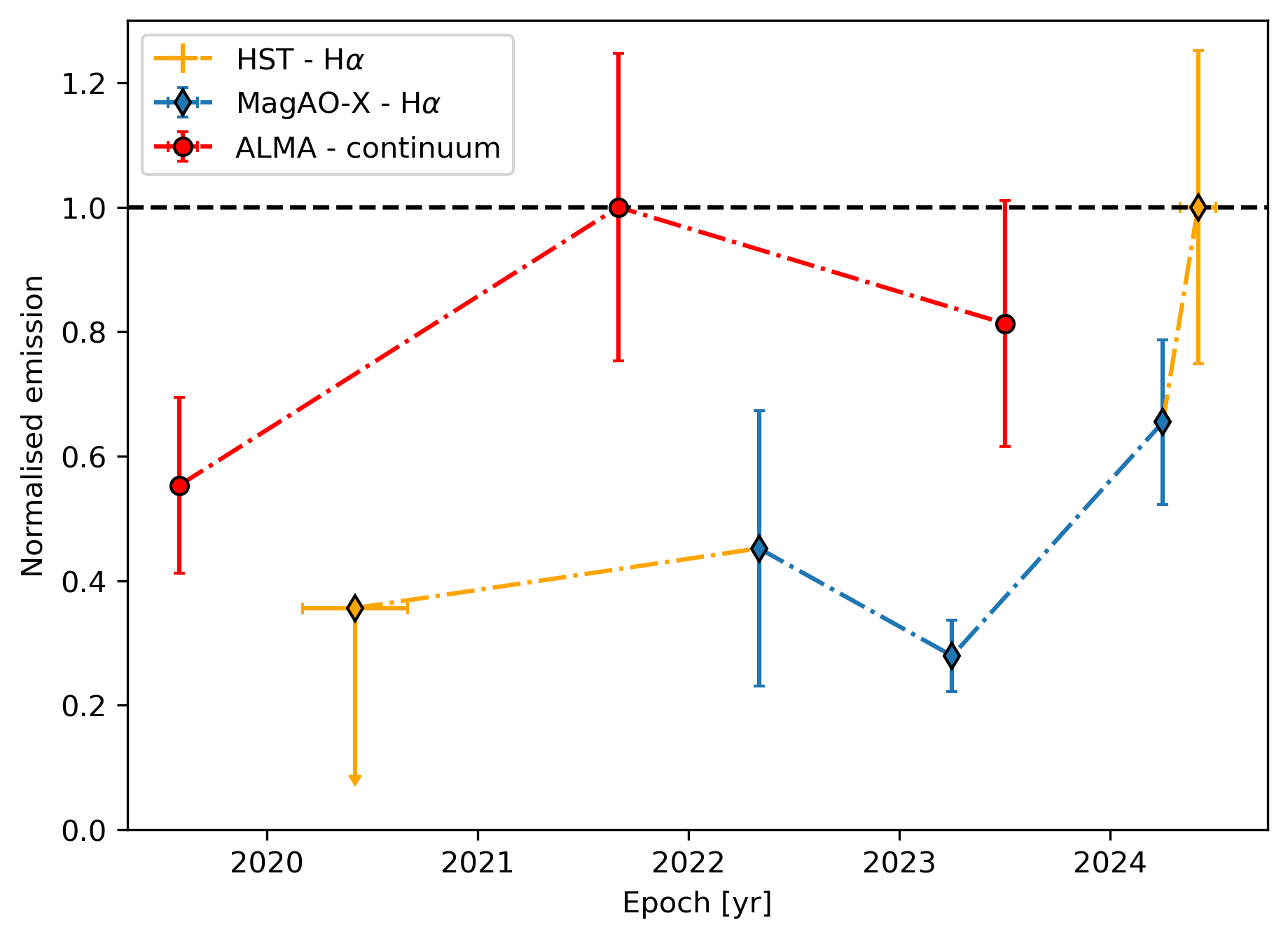}
      \caption{Comparison between the normalised H$\alpha$ fluxes of PDS~70 c observed with HST \citep[orange diamonds, ][]{Zhou_2021, Zhou_2025} and MagAO-X \citep[blue diamonds, ][]{Close_2025} and the normalised ALMA continuum peak flux of $c_{\rm smm}$ (red dots, this work). The black dashed line highlights the maximum of the normalisation.
              }
         \label{fig:Variability}
   \end{figure}

Recent multi-epoch H$\alpha$ detections of PDS 70 c using the Hubble Space Telescope (HST) \citep{Zhou_2021,Zhou_2025} and the MagAO-X instrument \citep{Close_2025} have shown evidence of variability. Indeed, these measurements have revealed an increase in flux of a factor $2.3$ between the 2023 and 2024 epochs observed with the MagAO-X instrument, and an additional $\sim40\%$ increase from the ground based detection to the most recent HST observation, taken a few weeks later. In Fig.~\ref{fig:Variability} we compare the normalised fluxes reported in \citep{Zhou_2021,Zhou_2025,Close_2025} with the normalised peak intensities of $c_{\rm smm}$ we measure in our ALMA images. A similar trend in the continuum emission is tentatively observed, although our measurements are statistically compatible with no significant variable emission (see Sec.~\ref{sec:Results}) and the lack of ALMA observation of the 2024 epoch prevents a proper comparison. The tentatively-inferred variable nature of the emission of $c_{\rm smm}$ can be due to a variety of physical mechanisms. Firstly, the mass advection from the protoplanetary disc to the circumplanetary environment can readily change its radius. A 40\% difference in radius can easily explain a factor of two difference in flux density, in particular for optically thick thermal emission. Secondly, within the assumption of thermal emission, the increase in peak intensity by a factor $\sim2$ that we measure with the ALMA data can be associated with an increase in temperature by the same factor as explained in \citet{Casassus_2022}. In the assumption that variability in the temperature structure is directly connected to the accretion luminosity onto the planet, we can outline the following argument. Considering the relation $L_{\rm acc}\propto T^4$ between temperature $T$ and accretion luminosity $L_{\rm acc}$, and that the accretion luminosity scales with the H$_\alpha$ luminosity with a power-law index of $\sim0.95$ for planetary mass accretion rates \citep{Aoyama_2021}, the sub-mm variability would imply a variability in H$_\alpha$ luminosity by a factor $\sim20$.  The measured H$_\alpha$ fluxes show an increase of $>3$, which is not as prominent. However, the lack of simultaneous ALMA and H$\alpha$ observations limits the interpretation of the results. Simultaneous observations of PDS 70 c tracing both H$_\alpha$ and ALMA continuum emission are needed in order to confirm a possible correlation between the variability of these two tracers.

\subsection{Spectral Index}

    \begin{table}
    \caption{Spectral index} 
    \label{table:Spectral Index} 
    \centering          
    \footnotesize
    \renewcommand{\arraystretch}{1.5}
    \begin{tabular}{c c c c}
    \hline\hline              lines
    & & \multicolumn{2}{c}{{ $\bf c_{\rm smm}$}}\\
    Dataset & $\nu_{\rm B7}$ & $\alpha_{\rm mm} \, (220 \rm GHz)$ & $\alpha_{\rm mm} \, (260 \rm GHz)$\\
     & [GHz] & & \\
    \hline                    
       B7 2019 & 350 & $1.2\pm0.8$ & $1.9\pm1.1$ \\
       B7 2021 & 344 & $2.6\pm0.8$ & $4.2\pm1.1$ \\
       B7 2023 & 344 & $2.1\pm0.8$ & $3.4\pm1.1$ \\
    \hline   
    \end{tabular}
    \end{table}

Observations at different frequencies of PDS 70 can be used to estimate a
spectral index. A detailed study of the spatial variations of the spectral index in PDS 70 was done in \citet{Doi_2024}, where Band 3 and 7 ALMA data were combined to suggest dust azimuthal and radial accumulation within the North West dust overdensity in the outer ring. In this work, since we can detect and resolve the emission from the inner disc, $c_{\rm smm}$ and the outer disc, we focus on the spectral index between Band 6 and Band 7.

Given two frequencies $\nu_{\rm B6}$ and $\nu_{\rm B7}$ with their associated flux density $F_{\rm B6}$ and $F_{\rm B7}$, the spectral index is given by

\begin{equation}
\label{eq:spix specific}
    \alpha_{\rm mm}=\frac{\log_{10}(F_{\rm B7}/F_{\rm B6})}{\log_{10}(\nu_{\rm B7}/\nu_{\rm B6})},
\end{equation} 
which can be generalised to 

\begin{equation}
\label{eq:spix general}
    \log_{10}F_\nu = \alpha_{\rm mm}\log_{10}\nu + \log_{10}F_0,
\end{equation}
where $F_0$ is a normalization factor. We calculate the spectral index in three different regions: the entire disc ($\alpha_{\rm mm}^{\rm D}$), the inner disc ($\alpha_{\rm mm}^{\rm ID}$) and $c_{\rm smm}$ ($\alpha_{\rm mm}^{c_{\rm smm}}$). For the first two regions, we consider the flux density from Table~\ref{table:Disc continuum properties} and Table~\ref{table:Inner disc continuum properties}. For the unresolved source $c_{\rm smm}$, instead, we use the task \texttt{imsmooth} of the CASA software to smooth our fiducial images with a Gaussian beam to the same resolution of the B6 2021 LF dataset\footnote{As the resolution in Band 6 is lower by a factor $\sim2$, we CLEAN the Band 7 datasets using an $uv$-taper to reach a similar resolution before applying the Gaussian smoothing.}. We measure a peak intensity of $c_{\rm smm}$ of $95\pm18~\rm\mu Jy/beam$, $172\pm30~\rm\mu Jy/beam$, $140\pm24~\rm\mu Jy/beam$, $54\pm17~\rm\mu Jy/beam$ and $55\pm14~\rm\mu Jy/beam$ for B7 2019, B7 2021, B7 2023, B6 2021 LF and B6 2021 HF, respectively, and use these estimates to compute the spectral index of $c_{\rm smm}$. Here we note that the peak emission we measure in these images is within $1\sigma$ of the values measured from the fiducial images (as expected for unresolved emission) with the exception of the Band 7 2021 dataset, for which we might suffer from contamination by the inner disc or diffuse emission. 

We then estimate the spectral index for each region using all the datasets simultaneously by fitting Eq.~\eqref{eq:spix general} with the \texttt{curve\_fit} function of the \texttt{scipy} module \citep{Virtanen_2020}, and we show the results of the fit in Fig.~\ref{fig:Spectral Index}. We obtain spectral indices of $\alpha_{\rm mm}^{\rm D}=2.8\pm0.4$, $\alpha_{\rm mm}^{c_{\rm smm}}=2.5\pm1.2$ and $\alpha_{\rm mm}^{\rm ID}=3.2\pm0.5$. The spectral index estimate for $c_{\rm smm}$ is affected by large uncertainties due to the emission variability in Band 7. As a result, for the circumplanetary material we also compute the spectral index for each epoch individually using Eq.~\eqref{eq:spix specific} and we report the results we obtain in Table~\ref{table:Spectral Index}.

Assuming that the emission is caused entirely by thermal emission from dust and neglecting self-scattering \citep{Zhu_2019}, there are two possible scenarios. If the dust is optically thick, the emission is characterised by a black body spectral index of $\alpha=2$ in the Rayleigh-Jeans limit. In the optically thin regime, instead, the spectral index will be given by $\alpha=\beta+2$, with $\beta$ depending on the properties of the dust grains \citep[see Fig. 4 and 10 from][]{Birnstiel_2018}. 

The spectral index of the inner disc $\alpha_{\rm mm}^{\rm ID}=3.2\pm0.5$ and the absence of variability in the single epoch estimates we measure indicate that the sub-mm emission is likely dominated by optically thin dust. 
By performing dust evolution and radiative transfer models of PDS 70, \citet{Pinilla_2024} found that the inner disc is long lived only if small grains (smaller than $0.1~\mu\rm m$) are diffused along with the gas into the gap carved by the two planets, in contrast with the larger grains trapped inside the ring. They computed the SED in the range $[0.45-7.50]~\rm mm$ and predicted spectral indices of $\alpha_{\rm mm}^{\rm D}=3.2$ and $\alpha_{\rm mm}^{\rm ID}=3.6$ for the entire disc and the inner disc, respectively, that are within $1\sigma$ error bars of our measurements. This filtration process showcases the impact that massive planets at large orbital separation can have on the terrestrial planet forming potential of inner discs.

\section{Conclusions}
\label{sec:Conclusions} 

In this paper we present new Band 7 and Band 6 high resolution observations of PDS 70. By performing 2D modelling of the dust emission in visibility space we:

\begin{itemize}
    \item Obtain a detailed model of the outer disc morphology of PDS 70. The residuals show the presence of complex substructures inside the region of the asymmetry.
    \item Re-detect the compact dust emission around PDS 70 c both in Band 7 ($345-355~\rm GHz$) with $>5\sigma$ and Band 6 ($220-260~\rm GHz$) with $>3\sigma$, confirming previous detections \citep{Isella2019, Benisty_2021, Casassus_2022} and marginally detect ($3\sigma$)  compact emission co-located with PDS 70 b in Band 6 for the first time.
    \item Measure an astrometric displacement of the material around PDS 70 c over three epochs consistent with the predicted positions of the planet \citep{Wang_2021}.
    \item Tentatively measure a peak intensity difference of up to $64\pm34\rm~\mu Jy/beam$ at $1\sigma$ confidence level for the region around PDS 70 c and perform a Bayesian test on our measurements, finding that they are not consistent with significant variable emission, while we do not find evidence of flux variability for the inner disc. 
    \item Observe variable morphology of the inner disc and resolve substructures for two epochs. We re-detect the clumps proposed in \citet{Casassus_2022} in the same dataset, but we do not detect them in the following epochs. However, after subtracting a Gaussian model for the inner disc emission, we do not detect any clump with more than $3\sigma$ significance. 
    \item Estimate the dust masses from the continuum integrated fluxes associated with both the emission around PDS 70 c and the inner disc. The results we obtain are consistent with previous estimates \citep{Benisty_2021}.
    \item Compute an integrated spectral index of $\alpha_{\rm mm}^{\rm ID}=3.2\pm0.5$, $\alpha_{\rm mm}^{\rm D}=2.8\pm0.4$, $\alpha_{\rm mm}^{c_{\rm smm}}=2.5\pm1.2$ for the inner disc, entire disc and the material detected around PDS 70 c, respectively. 
\end{itemize}

Observing at additional epochs with high resolution and SNR will help constrain the astrometry of the material around PDS 70 c more accurately and better understand the variability of the substructures in the inner disc. 

\begin{acknowledgements}
        We thank the anonymous referee for their feedback that helped improving this paper. MB acknowledges Marco Tazzari for the development of \texttt{Galario}. This project has received funding from the European Research Council (ERC) under the European Union’s Horizon 2020 research and innovation programme (PROTOPLANETS, grant agreement No. 101002188). S.F. acknowledges financial contributions by the European Union (ERC, UNVEIL, 101076613), and by PRIN-MUR 2022YP5ACE. L.R. is funded by the European Union (ERC, UNVEIL, 101076613). Views and opinions expressed are, however, those of the author(s) only and do not necessarily reflect those of the European Union or the European Research Council. Neither the European Union nor the granting authority can be held responsible for them. L.P. gratefully acknowledges support by the ANID BASAL project FB210003 and ANID FONDECYT Regular 1221442.  

        PC acknowledges support by the ANID BASAL project FB210003

        This work was supported by a Grant-in-Aid for JSPS Fellows, grant No. JP23KJ1008 (T.C.Y.).
        
        This paper makes use of the following ALMA data: ADS/JAO.ALMA\#2015.1.00888.S, ADS/JAO.ALMA\#2018.A.00030.S, ADS/JAO.ALMA\#2018.1.01774.S, ADS/JAO.ALMA\#2019.1.01138.S, ADS/JAO.ALMA\#2019.1.01619.S, ADS/JAO.ALMA\#2021.1.00782.S, ADS/JAO.ALMA\#2022.1.01695.S . ALMA is a partnership of ESO (representing its member states), NSF (USA) and NINS (Japan), together with NRC (Canada), NSTC and ASIAA (Taiwan), and KASI (Republic of Korea), in cooperation with the Republic of Chile. The Joint ALMA Observatory is operated by ESO, AUI/NRAO and NAOJ.

\end{acknowledgements}

\bibliographystyle{aa}
\bibliography{aa54959-25}

\begin{appendix}

\onecolumn

\section{Observations}\label{App: Observations}

\begin{table*}[h!]
\caption{Log of the ALMA observations presented in this work.}             
\label{table:observation_log}  
\centering                  
\footnotesize
\renewcommand{\arraystretch}{1.5}
\begin{tabular}{l c c c c c c r} 
\hline\hline           
Label & ID & Date & No. Ant. & Int. & Phase Cal. & Flux/Bandpass Cal. & PWV\\
 &  &  & &  [min] & &   & [mm] \\
\hline                  
B7 2021 & 2018.1.01774.S  & SB EB0: 18 July 2021  & 45  &   50   &  J1427-4206 & J1337-1257 & 0.93 \\
        & 2019.1.01138.S  & LB EB0: 6 August 2021  &  37  &  48   &  J1427-4206 & J1517-2422 &  0.40\\
\hline
B7 2023  & 2021.1.00782.S  & SB EB0: 2 March 2023 &  43  &  45  &  J1427-4206 &  J1256-0547 & 0.72 \\
  &    & SB EB1: 2 March 2023 &   43 &  45  &  J1427-4206 & J1256-0547 & 0.69  \\
  &  2022.1.01695.S  & SB EB0: 2 March 2023 &  43  & 34   & J1352-4412  & J1427-4206 &  0.77 \\
  &     & LB EB0: 1 June 2023 &  43  &  35  & J1352-4412  & J1427-4206 & 0.75 \\
  &     & LB EB1: 5 June 2023 &  44  &  35  & J1352-4412  & J1427-4206 & 0.66 \\
  &     & LB EB2: 5 June 2023 &  44  &  35  &  J1352-4412 & J1427-4206 & 0.65 \\
\hline
\end{tabular}
\tablefoot{Columns: Dataset label; ALMA Program ID; Observation date; Number of antennas used in the observation; Integration time; Phase calibrator; Flux and bandpass calibrator; Precipitable water vapour (PWV) level.}
\end{table*}

In Table~\ref{table:observation_log} we report the observing log associated to the ALMA datasets presented for the first time in this paper.

\section{Visibility Modelling}\label{App: Visibility Modelling}

    \begin{table*}[h!]
    \caption{\texttt{galario} best fit parameters}             
    \label{table:galario results}      
    \centering      
    \footnotesize
    \renewcommand{\arraystretch}{1.5}
    \begin{tabular}{c c c c c c c c c}     
    \hline\hline \\     
    Dataset & $\log_{10}f_1$ & $r_1$ & $\sigma_{r,1}$ & $\log_{10}f_2$ & $r_2$ & $\sigma_{r,2}$ & $\log_{10}f^a_1$ & $r^a_1$ \\ 
     & [Jy sr$^{-1}$] & [mas] & [mas] & [Jy sr$^{-1}$] & [mas] & [mas] & [Jy sr$^{-1}$] & [mas] \\ \\
    \hline\\                    
       B7 2019 & $10.0737_{-0.0008}^{+0.0008}$ & $658.2_{-0.4}^{+0.4}$ & $77.3_{-0.3}^{+0.3}$ & $9.703_{-0.003}^{+0.005}$ & $491.3_{-0.5}^{+0.6}$ & $38.2_{-0.6}^{+0.6}$ & $9.542_{-0.005}^{+0.008}$ & $636_{-2}^{+3}$\\  
       B7 2021 & $10.0887_{-0.0008}^{+0.0008}$ & $663.2_{-0.6}^{+0.6}$ & $79.7_{-0.4}^{+0.4}$ & $9.703_{-0.005}^{+0.005}$ & $497.4_{-0.7}^{+0.7}$ & $44.5_{-0.7}^{+0.7}$ & $9.66_{-0.02}^{+0.02}$ & $629\pm2$ \\
       B7 2023 & $10.1053_{-0.0006}^{+0.0005}$ & $666.5_{-0.2}^{+0.3}$ & $70.9_{-0.2}^{+0.2}$ & $9.806_{-0.002}^{+0.002}$ & $503.3_{-0.4}^{+0.5}$ & $46.6_{-0.4}^{+0.4}$ & $9.548_{-0.004}^{+0.005}$ & $628.5_{-1}^{+2}$ \\
       B6 2021 LF & $9.521_{-0.002}^{+0.002}$ & $682.7_{-0.2}^{+1.0}$ & $90.9_{-0.2}^{+0.7}$ & $8.836_{-0.002}^{+0.023}$ & $487.9_{-0.4}^{+2.3}$ & $55.7_{-0.3}^{+7.0}$ & $9.392_{-0.004}^{+0.006}$ & $701_{-1}^{+2}$ \\
       B6 2021 HF &  $9.69864_{-0.0008}^{+0.0008}$ & $643.7_{-0.3}^{+0.3}$ & $74.6_{-0.2}^{+0.3}$ & $9.228_{-0.004}^{+0.010}$ & $471.0_{-0.7}^{+0.6}$ & $33.3_{-1.0}^{+0.7}$ & $9.510_{-0.003}^{+0.005}$ & $650_{-1}^{+1}$ \\ \\
    \hline\hline \\          
    Dataset & $\sigma^a_{r,1}$ & $\theta^a_1$ & $\sigma^a_{\theta,1}$ & $i$ & $\rm PA$ & $\Delta \rm RA$ & $\Delta \rm Dec$ & \\ 
     & [mas] & [deg] & [deg] & [deg] & [deg] & [mas] & [mas] &  \\ \\
    \hline \\                   
       B7 2019  & $81_{-2}^{+2}$ & $142.51_{-0.18}^{+0.02}$ & $27.73_{-0.53}^{+0.01}$ & $49.94_{-0.01}^{+0.02}$ & $160.46_{-0.01}^{+0.06}$ & $-19.2_{-0.2}^{+0.2}$ & $2.3_{-0.2}^{+0.2}$ & \\  
       B7 2021 & $53_{-3}^{+3}$ & $138.3_{-0.5}^{+0.3}$ & $23.6_{-0.3}^{+0.3}$ & $49.83_{-0.02}^{+0.02}$ & $160.00_{-0.04}^{+0.05}$ & $-19.4_{-0.2}^{+0.2}$ & $2.4_{-0.2}^{+0.2}$ & \\
       B7 2023 & $72_{-10}^{+10}$ & $135.30_{-0.01}^{+0.15}$ & $28.10_{-0.45}^{+0.01}$ & $49.816_{-0.004}^{+0.010}$ & $160.40_{-0.02}^{+0.01}$ & $-18.6_{-0.1}^{+0.1}$ & $0.81_{-0.9}^{+0.9}$ & \\
       B6 2021 LF & $69_{-1}^{+2}$ & $140.07_{-0.01}^{+0.03}$ & $23.91_{-0.45}^{+0.04}$ & $49.983_{-0.004}^{+0.073}$ & $160.94_{-0.02}^{+0.04}$ & $-5.5_{-0.1}^{+0.3}$ & $2.4_{-0.1}^{+0.3}$ & \\
       B6 2021 HF & $60.3_{-0.7}^{+0.6}$ & $139.53_{-0.18}^{+0.01}$ & $23.82_{-0.19}^{+0.01}$ & $50.097_{-0.005}^{+0.021}$ & $160.64_{-0.01}^{+0.04}$ & $-14.4_{-0.2}^{+0.2}$ & $2.2_{-0.2}^{+0.2}$ & \\ \\
    \hline
    \end{tabular}
    \tablefoot{In the B6 2021 HF model, we enforce $r_1>r_2$.}
    \end{table*}

    In Table~\ref{table:galario results} we summarise the best fit parameters resulting from our visibility modelling of PDS 70 Band 7 and 6 observations. We have adopted the following priors: $\log_{10}f_1=[8.0, 11.0]$, $r_1=[0, 1.0]\rm~arcsec$, $\sigma_{r,1}=[0, 1.0]\rm~arcsec$, $\log_{10}f_2=[8.0, 11.0]$,  $r_2=[0, 1.0]\rm~arcsec$, $\sigma_{r,2}=[0, 1.0]\rm~arcsec$, $\log_{10}f^{\rm a}_1=[8.0, 11.0]$,  $r^{\rm a}_1=[0, 1.0]\rm~arcsec$, $\sigma^{\rm a}_{r,1}=[0, 1.0]\rm~arcsec$, $\theta^{\rm a}_1=[120, 180]\rm~arcsec$, $\sigma^{\rm a}_{\theta,1}=[0, 45]\rm~arcsec$, $i=[0, 90]\rm~deg$, $\rm PA=[0, 180]\rm~deg$, $\rm\Delta Ra=[-2.0, 2.0]\rm~arcsec$, $\rm\Delta Dec=[-2.0, 2.0]\rm~arcsec$. For the test run, we additionally considered the following priors for the inner Gaussian: $\log_{10}f_0=[8.0, 11.0]$, $\sigma_{r,0}=[0, 1.0]\rm~arcsec$ 

\section{Inner Disc Clumps}\label{App: Inner Disc Clumps}

 \begin{figure*}[!h]
    \centering
    \includegraphics[width=0.8\textwidth]{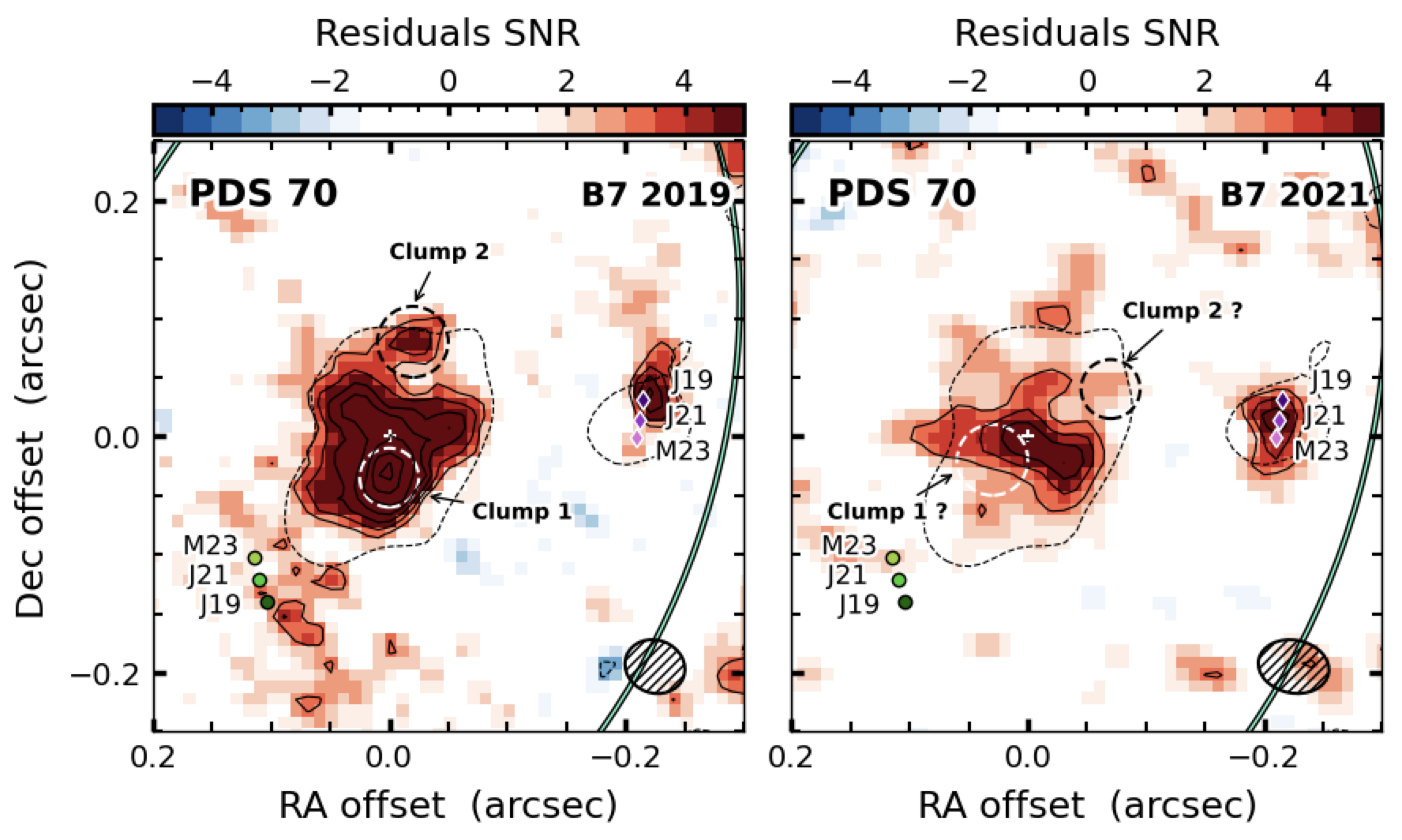}
    
 \caption{Same as left an middle panels of Fig.~\ref{fig:Residual cavity all B7}, with additional labels to identify Clump 1 and Clump 2. The dashed white and black circles represent the location where we re-detect Clump 1 and Clump 2, respectively, in the B7 2019 dataset (left panel), and their expected position assuming Keplerian rotation in the B7 2021 dataset (right panel).}
 \label{fig:Inner disc clumps}
    \end{figure*}

In Fig.~\ref{fig:Inner disc clumps} we show the left and middle panels of Fig.~\ref{fig:Residual cavity all B7}, with additional labels to point out the position of the clumps proposed in \citet{Casassus_2022}. We show the location where we re-detect Clump 1 and Clump 2 in the B7 2019 dataset using white and black dashed circles, respectively. For the B7 2021 dataset, we use the same circles to show the regions where we expect to detect the two clumps assuming Keplerian rotation. In the region associated with Clump 2 we tentatively see the presence of compact emission with $<3\sigma$ significance, although it is dominated by noise due to the low SNR of this dataset, whether we do not detect any clump in the region associated with Clump 1.  

\section{Spectral Index}\label{App: Spectral Index}

    \begin{figure*}[h!]
    \centering 
        \begin{subfigure}[t]{0.3\columnwidth}
        \centering
        \includegraphics[width=\columnwidth]{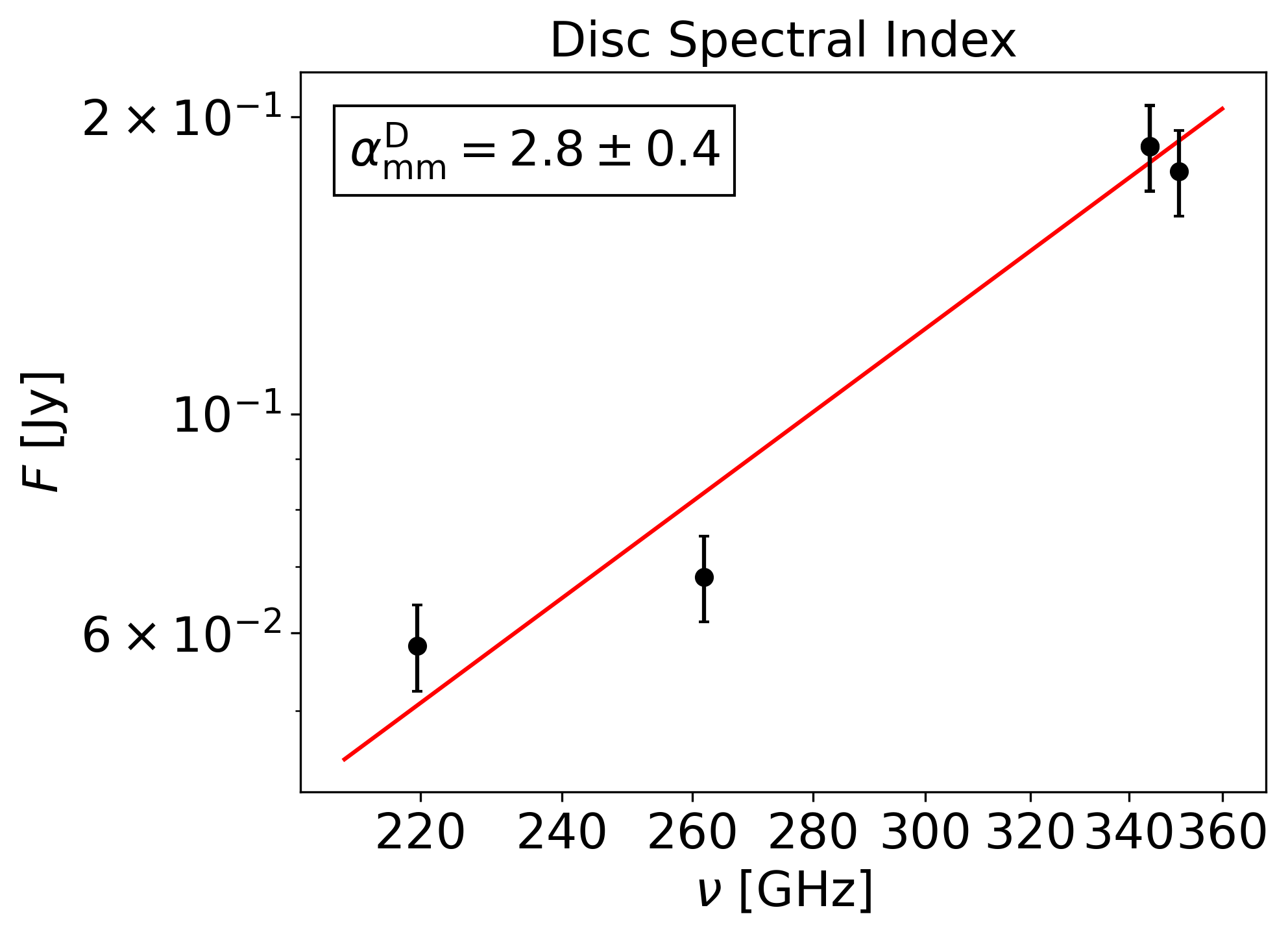}
        \caption{
              }
         \label{fig:Disc Spectral Index}
         \end{subfigure}
         \hfill
        \begin{subfigure}[t]{0.3\columnwidth}
        \centering
        \includegraphics[width=\columnwidth]{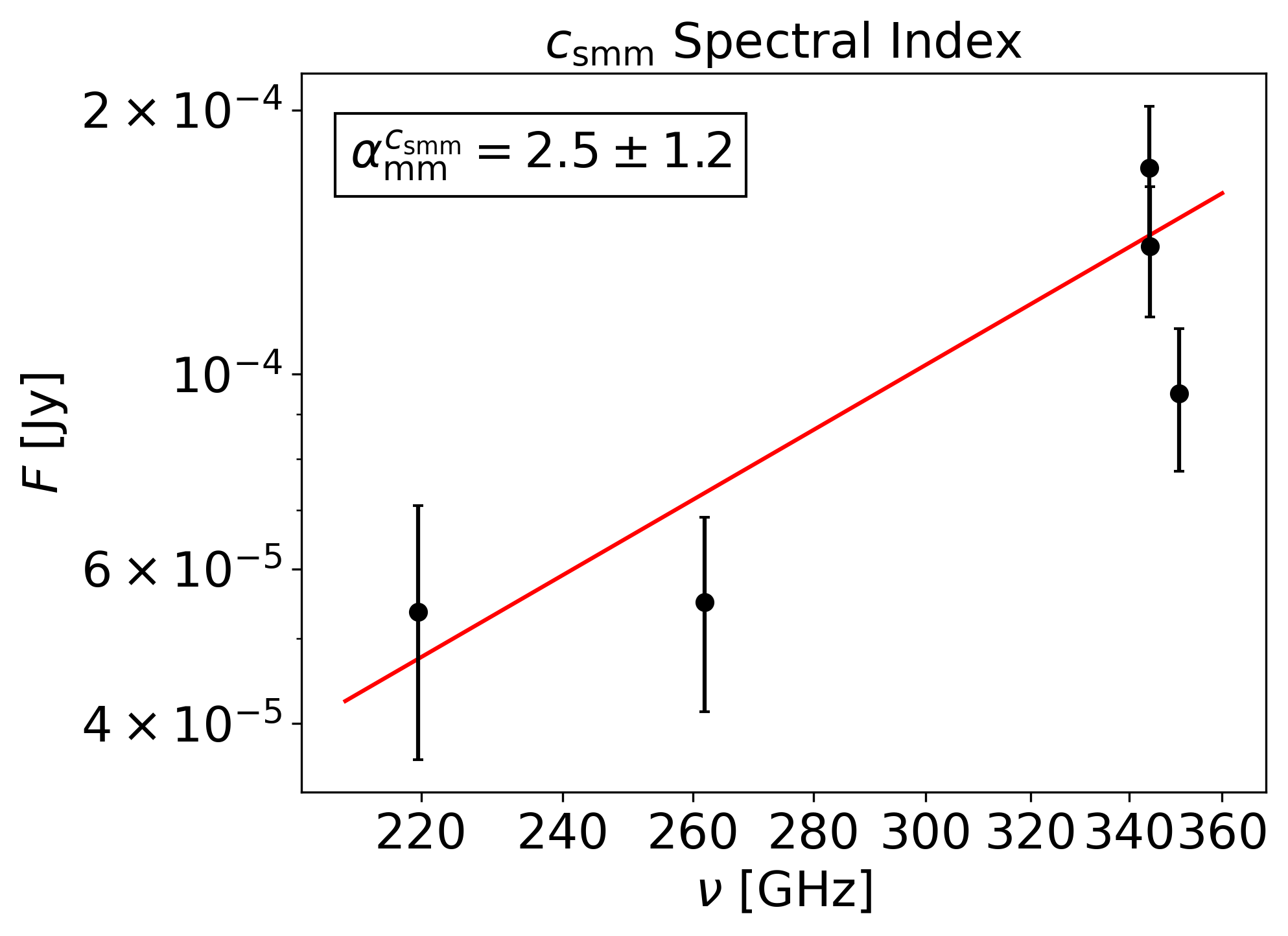}
        \caption{
              }
         \label{fig:CPD Spectral Index}
        \end{subfigure}
        \hfill 
        \begin{subfigure}[t]{0.3\columnwidth}
        \centering
        \includegraphics[width=\columnwidth]{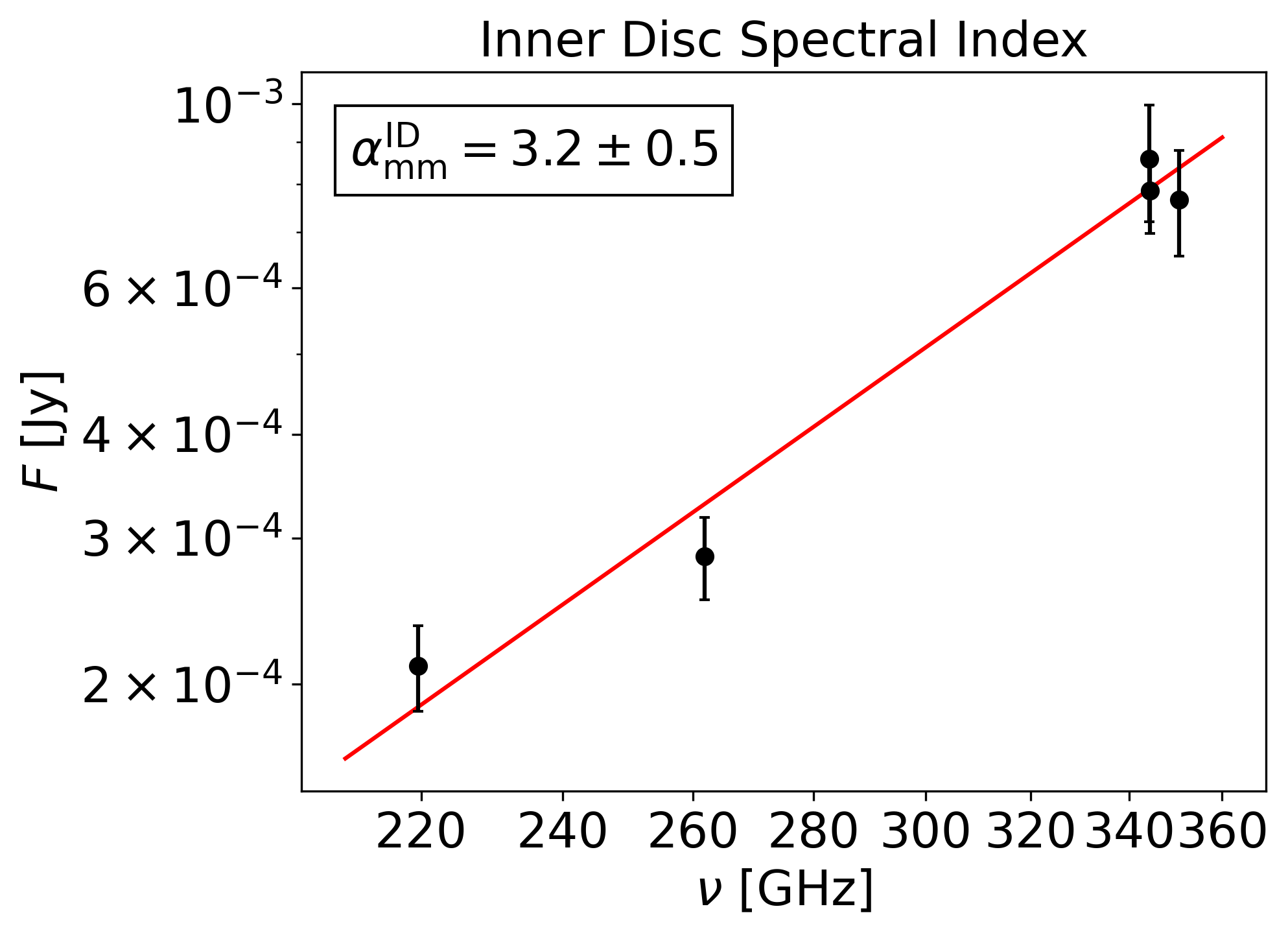}
        \caption{
              }
         \label{fig:Inner Disc Spectral Index}
         \end{subfigure}
    \caption{SED for the Band 6 and 7 fluxes. a) SED of the full disc emission. b) SED of the $c_{\rm smm}$ emission. c) SED of the inner disc emission around PDS 70 c. The black dots represent the measured fluxes reported in Tables~\ref{table:CPD continuum properties}-\ref{table:Inner disc continuum properties}, while the red line represent the best fit for the spectral index. 
              }
    \label{fig:Spectral Index}
   \end{figure*}

In Fig.~\ref{fig:Spectral Index} we show the Spectral Energy Distribution (SED) of the CPD and inner disc of PDS 70 using the datasets considered in this Article and we compare it with the best fit model of the spectral index.  

\end{appendix}

\end{document}